# Audio Watermarking with Error Correction

Aman Chadha, Sandeep Gangundi, Rishabh Goel, Hiren Dave, M. Mani Roja
Department of Electronics & Telecommunication
Thadomal Shahani Engineering College
Mumbai, INDIA

*Abstract*— In recent times, communication through the internet has tremendously facilitated the distribution of multimedia data. Although this is indubitably a boon, one of its repercussions is that it has also given impetus to the notorious issue of online music piracy. Unethical attempts can also be made to deliberately alter such copyrighted data and thus, misuse it. Copyright violation by means of unauthorized distribution, as well as unauthorized tampering of copyrighted audio data is an important technological and research issue. Audio watermarking has been proposed as a solution to tackle this issue. The main purpose of audio watermarking is to protect against possible threats to the audio data and in case of copyright violation or unauthorized tampering, authenticity of such data can be disputed by virtue of audio watermarking.

*Keywords- watermarking; audio watermarking; data hiding; data confidentiality.*

## I. INTRODUCTION

Over the years, there has been tremendous growth in computer networks and more specifically, the internet. This phenomenon, coupled with the exponential increase of computer performance, has facilitated the distribution of multimedia data such as images, audio, video etc. Data transmission has been made very simple, fast and accurate using the internet. However, one of the main problems associated with transmission of data over the internet is that it may pose a security threat, i.e., personal or confidential data can be stolen or hacked in many ways. Publishers and artists, hence, may be reluctant to distribute data over the Internet due to lack of security; copyrighted material can be easily duplicated and distributed without the owner's consent. Therefore, it becomes very important to take data security into consideration, as it is one of the essential factors that need attention during the process of data distribution. Watermarks have been proposed as a way to tackle this tough issue. This digital signature could discourage copyright violation, and may help determine the authenticity and ownership of an image.

Watermarking is "the practice of imperceptibly altering a Work to embed a message about that Work" [1]. Watermarking can be used to secretly transmit confidential messages, for e.g., military maps, without the fact of such transmission being discovered. Watermarking, being ideally imperceptible, can be essentially used to mask the very existence of the secret message [6]. In this manner, watermarking is used to create a covert channel to transmit confidential information [5]. Watermarking is an effective means of hiding data, thereby protecting the data from unauthorized or unwanted viewing.

Watermarking is becoming increasingly popular, especially for insertion of undetectable identifying marks, such as author or copyright information to the host signal. Watermarking may probably be best used in conjunction with another data-hiding method such as steganography, cryptography etc. Such data-hiding schemes, when coupled with watermarking, can be a part of an extensive layered security approach.

To combat online music piracy, a digital watermark could be added to all recordings prior to release, signifying not only the author of the work, but also the user who has purchased a legitimate copy. Audio watermarking is defined as "the imperceptible, robust and secure communication of data related to the host audio signal, which includes embedding into, and extraction from, the host audio signal" [4]. Digital audio watermarking involves the concealment of data within a discrete audio file. Intellectual property protection is currently the main driving force behind research in this area. Several other applications of audio watermarking such as copyright protection, owner identification, tampering detection, fingerprinting, copy and access control, annotation, and secret communication, are in practice. Other related uses for watermarking include embedding auxiliary information which is related to a particular song, like lyrics, album information, or a hyperlink etc. Watermarking could be used in voice conferencing systems to indicate to others which party is currently speaking. A video application of this technology would consist of embedding subtitles or closed captioning information as a watermark [7].

## II. IDEA OF THE PROPOSED SOLUTION

### A. Process of Watermarking

The block diagram for watermarking is as shown below:

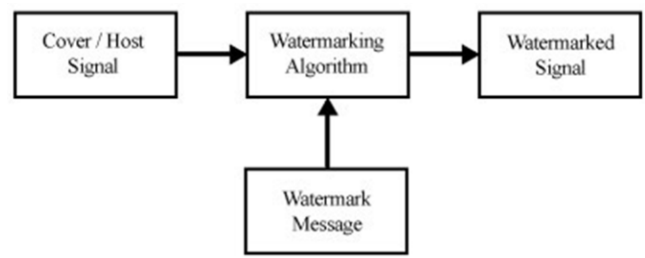

Figure 1. General watermarking block diagram

We can summarize the entire process of hiding and retrieving data as follows:

- Read the data to be hidden.
- Read the cover, i.e., host, in which data is to be hidden.
- Apply watermarking methods on the host.
- Hide the data in the host.
- Retrieve the original data at the receiver end.

*B. Mean Square Error*

Mean Square Error (MSE) [2], first introduced by C. F. Gauss, serves as an important parameter in gauging the performance of the watermarking system. The following factors justify the choice of MSE as a convenient and extensive standard for performance assessment of various techniques of audio watermarking:

*1) Simplicity:* It is parameter-free and inexpensive to compute, with a complexity of only one multiply and two additions per sample. It is also memory less, i.e., MSE can be evaluated at each sample, independent of other samples.

*2) Clear physical meaning:* It is the natural way to define the energy of the error signal. Such an energy measure is preserved even after any orthogonal or unitary linear transformation. The energy preserving property guarantees that the energy of a signal distortion in the transform domain is the same as that in the signal domain.

*3) Excellent metric in the context of optimization:* The MSE possesses the properties of convexity, symmetry, and differentiability.

*4) Used as a convention:* It has been extensively employed for optimizing and assessing a wide variety of signal processing applications, including filter design, signal compression, restoration, reconstruction, and classification.

MSE is essentially a signal fidelity measure [14],[15]. The goal of a signal fidelity measure is to compare two signals by providing a quantitative score that describes the degree of similarity/fidelity or, conversely, the level of error/distortion between them. Usually, it is assumed that one of the signals is a pristine original, while the other is distorted or contaminated by errors.

Suppose that $x = \{x_i \mid i = 1, 2, \ldots, N\}$ and $y = \{y_i \mid i = 1, 2, \ldots, N\}$ are two finite-length, discrete signals, for e.g., visual images or audio signals. The MSE between the signals is given by the following formula:

$$MSE(x, y) = \frac{1}{N} \sum_{i=1}^{N} (x_i - y_i)^2 \qquad (1)$$

Where,

N is the number of signal samples.
$x_i$ is the value of the $i^{th}$ sample in x.
$y_i$ is the value of the $i^{th}$ sample in y.

### III. IMPLEMENTATION STEPS

Audio Watermarking can be implemented in 3 ways:

- Audio in Audio
- Audio in Image
- Image in Audio

*A. Audio in Audio*

In this method, both the cover file and the watermark file are audio signals. The watermark signal must have fewer samples as compared to those of the cover audio signal. Further, this method can be implemented with the help of two techniques, namely, Interleaving and DCT.

*1) Using Interleaving:* It is a way to arrange data in a non-contiguous way so as to increase performance [3]. The following example illustrates the process of interleaving:

Original signal:           AAAABBBBCCCCDDDDEEEE
Interleaved signal:       ABCDEABCDEABCDEABCDE

In this technique, the samples of watermark audio are inserted in between the samples of the cover audio file [9],[10]. In terms of complexity, this is the simplest method of audio watermarking.

*2) Using Discrete Cosine Transform:* This technique is based on the Discrete Cosine Transform (DCT) [11]-[13]. In this technique, we take the DCT of both the cover audio and the watermark audio signals. Upon zigzag scanning, the high frequency DCT coefficients of the cover audio file are replaced with the low frequency DCT coefficients of the watermark audio file. During transmission, the Inverse Discrete Cosine Transform (IDCT) of the final watermarked DCT is taken. In this particular technique, since both the host, i.e., cover and watermark signal are in the audio format, we implement this method using a 1D DCT which is defined by the following equation:

$$F(u) = \sqrt{\frac{2}{N}} \sum_{i=0}^{N-1} A(i) * \cos\left(\frac{u(2i+1)\pi}{2N}\right) * f(i) \qquad (2)$$

Where,

$$A(i) = \begin{cases} \frac{1}{\sqrt{2}} & \text{for } u = 0 \\ 1 & \text{otherwise} \end{cases}$$

f(i) is the input sequence.

*B. Audio in Image*

This watermarking implementation uses DCT for embedding audio file in an image. Here we take DCT of both the cover image and the watermark audio files. The low frequency coefficients of both the DCT's are taken. The high frequency coefficients of the DCT of the image are replaced with the low frequency coefficients of the DCT of the watermark audio file. During transmission, the IDCT of the final watermarked DCT is taken. This technique involves both an audio signal (watermark) and an image (host). Hence, we implement this method using a 1D DCT for the audio signal and a 2D DCT for the image. Equation (2) defines a 1D DCT while the corresponding equation for a 2D DCT is defined by the following equation:

$$F(u,v) = \sqrt{\frac{2}{N}}\sqrt{\frac{2}{M}}\sum_{i=0}^{N-1} A(i) * \cos\left(\frac{u(2i+1)\pi}{2N}\right) * \sum_{j=0}^{M-1} A(j) * \cos\left(\frac{v(2j+1)\pi}{2M}\right) * f(i,j) \quad (3)$$

Where,

$$A(i) = \begin{cases} \frac{1}{\sqrt{2}} & \text{for } u = 0 \\ 1 & \text{otherwise} \end{cases}$$

$$A(j) = \begin{cases} \frac{1}{\sqrt{2}} & \text{for } v = 0 \\ 1 & \text{otherwise} \end{cases}$$

f(i, j) is the 2D input sequence.

*C. Image in Audio*

In this implementation, as deployed previously, DCT is used for embedding an image in an audio file. Here we take the DCT of both the cover audio and the watermark image files. This is followed by zigzag scanning so as to ascertain the low frequency and high frequency DCT coefficients. The high frequency DCT coefficients of the audio signal are replaced with the low frequency DCT coefficients of the watermark image file. While transmitting, the IDCT of the final watermarked DCT is taken.

Since this method is similar to the 'Audio in Image' technique with respect to the parameters involved, i.e., this technique also involves both an audio signal (watermark) and an image (host), hence, we may implement this method using a 1D DCT for the audio signal and a 2D DCT for the image. Equation (2) defines a 1D DCT while, (3) defines a 2D DCT.

IV. RESULTS

*A. Audio in Audio*

The spectra of the input, i.e., original cover and watermark audio signal are as shown below:

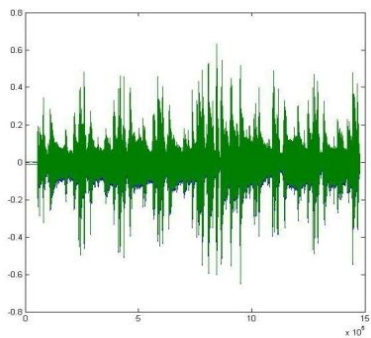

Figure 2. Original cover audio signal

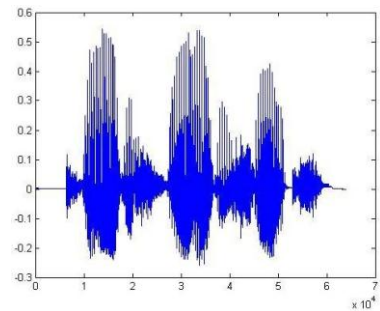

Figure 3. Original watermark audio signal

*1) Using Interleaving:* The spectra of the output, i.e., watermarked audio signal and the recovered audio signal, obtained by interleaving are as shown below:

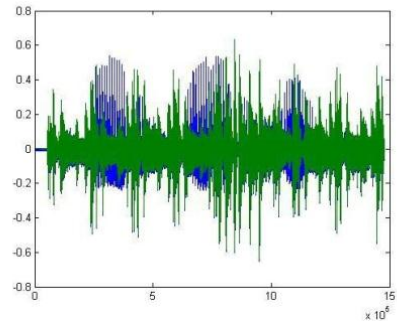

Figure 4. Watermarked audio signal

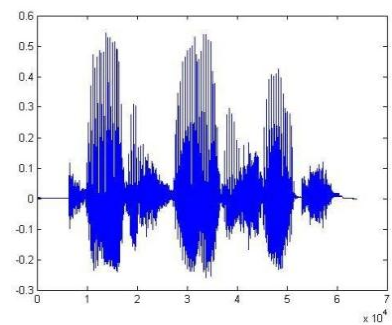

Figure 5. Recovered audio watermark signal

*2) Using DCT:* The spectra of the output, i.e., watermarked audio signal and the recovered audio signal, obtained by DCT based 'Audio in Audio' watermarking are as shown below:

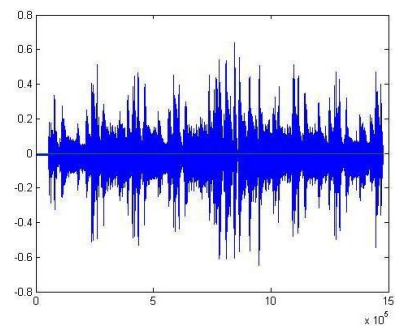

Figure 6. Watermarked audio signal

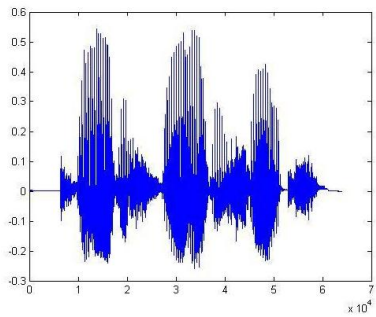
Figure 7. Recovered audio watermark signal

*B. Audio in Image*

The input, i.e., original cover image and spectrum of the watermark audio signal is as shown below:

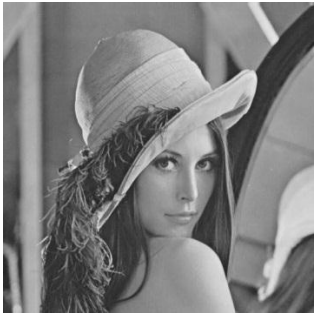
Figure 8. Original cover image

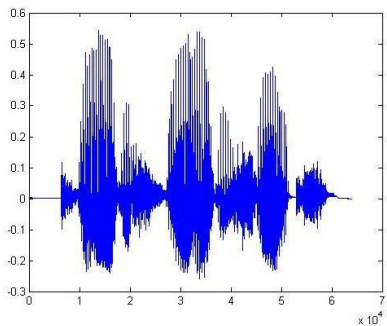
Figure 9. Original watermark audio signal

The output, i.e., watermarked image and the spectrum of the recovered audio signal, obtained by 'Audio in Image' watermarking is as shown below:

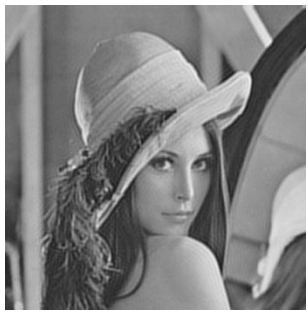
Figure 10. Watermarked image

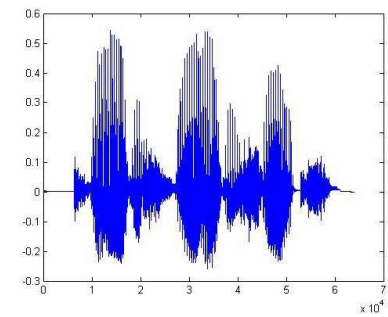
Figure 11. Recovered audio watermark signal

*C. Image in Audio*

The input, i.e., spectrum of the original cover signal and the watermark image is as shown below:

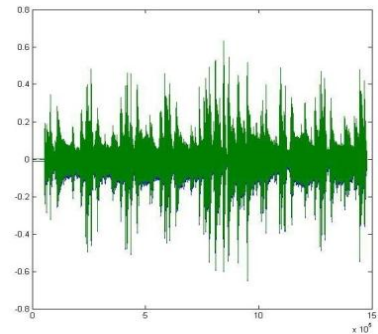
Figure 12. Original cover audio signal

**CODE IS 89ASDF**
Figure 13. Original watermark image

The output, i.e., spectrum of the watermarked signal and the recovered watermark image, obtained by 'Image in Audio' watermarking is as shown below:

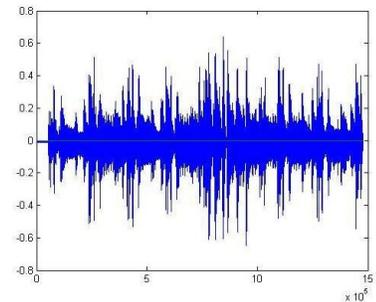
Figure 14. Watermarked image

**CODE IS 89ASDF**
Figure 15. Recovered watermark

For all practical purposes, it is preferable to have a quantitative measurement to provide an objective judgment of the extracting fidelity. This is done by calculating the MSE in each case. Results, thus obtained, have been tabulated.

TABLE I. CALCULATION OF MSE WITHOUT NOISE

| Types | MSE | |
|---|---|---|
| | *Watermarked signal* | *Watermark* |
| Audio in Audio (Interleaving) | $3.5 \times 10^{-4}$ | 0 |
| Audio in Audio (DCT) | $3.4 \times 10^{-3}$ | $2.48 \times 10^{-9}$ |
| Image in Audio | $3.2 \times 10^{-3}$ | 0 |
| Audio in Image | 23.3 | $2.47 \times 10^{-9}$ |

## V. ERROR CORRECTION USING HAMMING CODES

During transmission, when data travels through a wireless medium over a long distance, it is highly probable that it may get corrupted due to fluctuations in channel characteristics or other external parameters. Hence, hamming codes, which are Forward Error Correction (FEC) codes, may be used for the purpose of error correction in audio watermarking.

Hamming codes can detect up to two simultaneous bit errors and correct single bit errors; thus, reliable communication is possible when the "hamming distance" between the transmitted and received bit patterns is less than or equal to one [8]. In contrast, simple parity codes cannot correct errors and can only detect an odd number of errors. In this application, we are using the (15, 11) form of the hamming code. The improvement in the quality of received signal is clearly visible from the following waveforms:

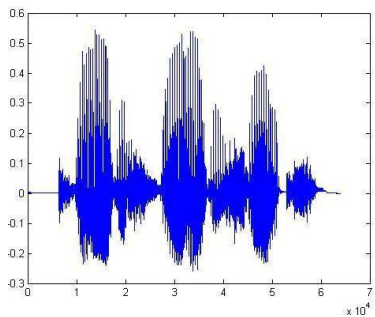

Figure 16. Desired audio signal

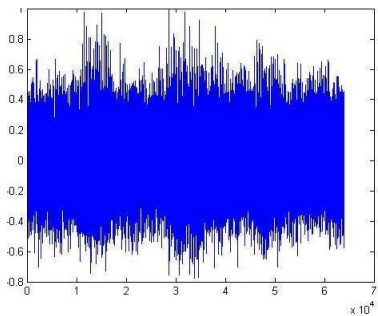

Figure 17. Recovered audio signal without Hamming Code

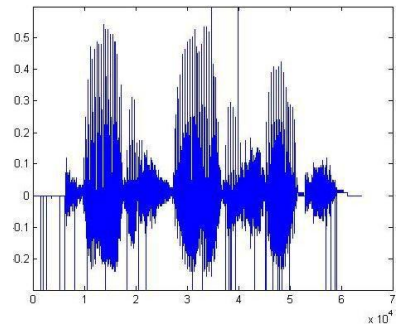

Figure 18. Recovered audio signal with Hamming Code

The corresponding results for Audio Watermarking with noise have been tabulated as follows:

TABLE II. CALCULATION OF MSE WITH NOISE

| Types | MSE | |
|---|---|---|
| | *Watermarked with noise* | *Watermark with noise* |
| Audio in Audio (Interleaving) | 0.010 | 0.010 |
| Audio in Audio (DCT) | 0.002 | 0.008 |
| Image in Audio | 0.003 | $1.92 \times 10^3$ |
| Audio in Image | 323.3 | 0 |

## VI. CONCLUSION

DCT is an effective and robust algorithm for audio watermarking as the audio signal retrieved is clearly audible. Hamming Codes enable data correction in case of data corruption during transmission and help in recovering the original audio signal.